\documentclass[sn-aps]{sn-jnl}

\usepackage{graphicx}%
\usepackage{multirow}%
\usepackage{amsmath,amssymb,amsfonts}%
\usepackage{amsthm}%
\usepackage{mathrsfs}%
\usepackage[title]{appendix}%
\usepackage{xcolor}%
\usepackage{textcomp}%
\usepackage{manyfoot}%
\usepackage{booktabs}%
\usepackage{algorithm}%
\usepackage{algorithmicx}%
\usepackage{algpseudocode}%
\usepackage{listings}%
\usepackage{natbib}
\UseRawInputEncoding    

\usepackage[justification=centering]{caption}   

\begin{document}

\title[Article Title]{Fabrication and Characterization of X-ray TES Detectors Based on Annular AlMn Alloy Films}


\author*[1]{\fnm{Yifei} \sur{Zhang}}\email{zhangyf@ihep.ac.cn}
\author[1]{\fnm{Zhengwei} \sur{Li}}
\author[2]{\fnm{Mengxian} \sur{Zhang}}
\author[1]{\fnm{Guofu} \sur{Liao}}
\author[1]{\fnm{Zhouhui} \sur{Liu}}
\author[1]{\fnm{Yu} \sur{Xu}}
\author[3]{\fnm{Nan} \sur{Li}}
\author[4]{\fnm{Liangpeng} \sur{Xie}}
\author[4]{\fnm{Junjie} \sur{Zhou}}
\author[1]{\fnm{Xufang} \sur{Li}}
\author[1]{\fnm{He} \sur{Gao}}
\author[1]{\fnm{Shibo} \sur{Shu}}
\author[1]{\fnm{Yongping} \sur{Li}}
\author[1]{\fnm{Yudong} \sur{Gu}}
\author[1]{\fnm{Daikang} \sur{Yan}}
\author[1]{\fnm{Xuefeng} \sur{Lu}}
\author[1]{\fnm{Hua} \sur{Feng}}
\author[1]{\fnm{Yongjie} \sur{Zhang}}
\author*[1]{\fnm{Congzhan} \sur{Liu}}\email{liucz@ihep.ac.cn}

\affil*[1]{\orgname{Key Laboratory of Particle Astrophysics, Institute of High Energy Physics, CAS}, \orgaddress{\state{Beijing}, \postcode{100049}, \country{China}}}

\affil[2]{\orgdiv{School of Mechanical and Elecrical Engineering}, \orgname{China University of Mining and Technology}, \orgaddress{\city{Xuzhou}, \postcode{221116}, \country{China}}}

\affil[3]{\orgname{Institute of Frontier and Interdisciplinary Science, Shandong University}, \orgaddress{\city{Qingdao}, \postcode{266237}, \country{China}}}

\affil[4]{\orgname{School of Physics and Materials, Nanchang University}, \orgaddress{\city{Nanchang}, \postcode{330031}, \country{China}}}


\abstract{AlMn alloy films are widely fabricated into superconducting transition edge sensors (TESs) for the detection of cosmic microwave background radiation. However, the application in X-ray or gamma-ray detection based on AlMn TES is rarely reported. In this study, X-ray TES detectors based on unique annular AlMn films are developed. The fabrication processes of TES detectors are introduced in detail. The characteristics of three TES samples are evaluated in a dilution refrigerator. The results demonstrate that the I-V characteristics of the three annular TES detectors are highly consistent. The TES detector with the smallest absorber achieved the best energy resolution of 11.0 eV @ 5.9 keV, which is inferior to the theoretical value. The discrepancy is mainly attributed to the larger readout electronics noise than expected.}




\maketitle

\section{Introduction}\label{sec1}

The superconducting Transition-Edge Sensor (TES) has been widely used in the detections from millimeter wave to gamma ray due to its extremely low equivalent power or high energy resolution \citep{Irwin-2005}. For X-ray photon detection, the best energy resolution reaches 1.6 eV @ 5.9 keV \citep{Smith-2012}, which is nearly two orders of magnitude better than that of semiconductor detector. It has become a key detector to solve frontier physics problems such as searching for the missing baryon in the warm-hot intergalatic medium and distinguishing the chemical state in different compounds \citep{Uhlig-2015, Doriese-2017, Pajot-2018, Cui-2020}. In X-ray detection applications, Ti/Au and Mo/Au bilayer films have been widely used, while alloy films such as AlMn films developed simultaneously over twenty years ago are rarely reported. The main reason is that it is inconvenient to adjust the critical temperature of alloy film by doping \citep{Young-2004, Jalkanen-2007}. This situation remained unchanged until a baking-based Tc adjusting method was developed \citep{Li-2016}. This method can easily change the critical temperature of AlMn films across a range spanning tens to hundreds of millikelvins. Several experiments have confirmed this approach \citep{Vavagiakis-2019, Lv-2020, Yu-2021}. In addition, it is easier to prepare AlMn film than bilayer film by magnetron sputtering. It has also been demonstrated that the AlMn alloy film is less affected by the magnetic field \citep{Vavagiakis-2018}. These advantages have already resulted in the AlMn film being widely applied in some fields. It has been used by several Cosmic Microwave Background (CMB) detection projects, such as SPT-3G \citep{Vavagiakis-2018}, POLARBEAR-2 \citep{Westbrook-2018}, Simons Observatory \citep{Stevens-2020}, and AliCPT-1 \citep{Salatino-2021}. Wang used AlMn TES with low $T_c$ in RICOCHET experiment to measure the coherent elastic neutrino nucleus scattering signals \citep{Wang-2024}. Lv developed a X-ray calorimeter with the AlMn TES and got a resolution of 17.5 eV @ 5.9 keV \citep{Lv-2020-thesis}. Here, we study the technology of AlMn TES detectors and hope to put them on future X-ray or gamma-ray astronomy satellites (such as the Wide X-ray Polarization Telescope \citep{Yin-2022}, and another concept mission to investigate the origin of 511 keV lines at the Galactic Center) for scientific observation. 

AlMn alloy films exhibit relatively higher resistivity than commonly used bilayer films such as Mo/Au or Mo/Cu. To achieve lower resistance, AlMn TESs employed in CMB detection are usually designed as elongated rectangle with a high width-to-length ratio \citep{Vavagiakis-2018, Westbrook-2018, Stevens-2020, Salatino-2021}. However, when a narrow rectangle TES is used in X-ray or gamma-ray detection and operated at the recommended temperature of around 100 mK, the thermal conductivity, which is proportional to the TES perimeter, becomes constrained by the fixed-area suspended membrane. One solution to solve this issue involves incorporating auxiliary structures, such as copper bars, to effectively increase the perimeter \citep{Hays-Wehle-2016}. Adding finger-like extensions to a squre TES can also reduce resistance, though this approach has not yet been explored for AlMn TESs. In this work, we propose a novel annular-shaped TES design as an alternative to the conventional rectangular geometry. This innovative structure enables control over the normal resistance by tuning the inner-to-outer radius ratio. Meanwhile, the outer radius can extend to the edge of the suspension film, enabling significant adjustability of thermal conductivity. Basing on this design, we fabricated annular AlMn X-ray detectors and characterized them through I-V tests. The detectors were calibrated using 5.9 keV X-ray photons generated by a $^{55}$Fe radioactive source.

\section{Detector design and fabrication}\label{sec2}

\begin{figure}[h]
    \centering
    \includegraphics[width=0.7\textwidth]{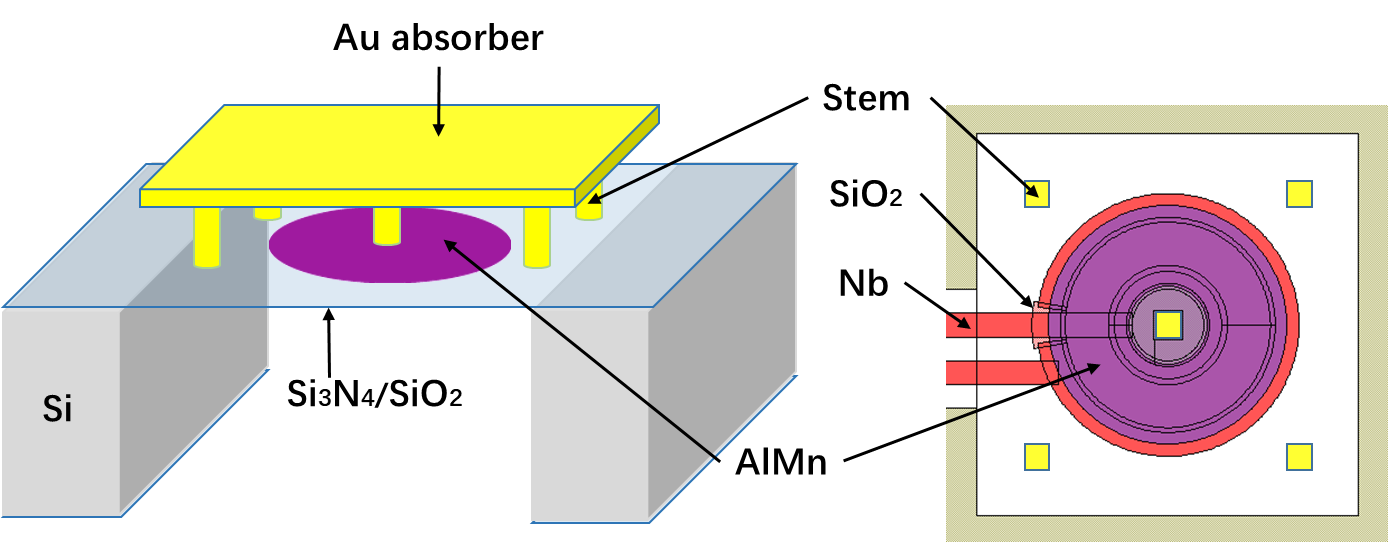}
    \caption{Structure diagram of TES detector. The left panel gives a three-dimensional schematic picture. The right panel is a top view of the TES detector without absorber.}
    \label{fig1}
\end{figure}

This study utilized a 4-inch silicon wafer substrate composing a 200 $\mu$m thick silicon core sandwiched between 1 $\mu$m Si$_3$N$_4$ and 0.2 $\mu$m SiO$_2$ layers on both sides. An AlMn TES detector, as shown in Fig. \ref{fig1}, was suspended on the Si$_3$N$_4$ membrane. The core structure consisted of a gold (Au) absorber, an AlMn alloy film coupled with two niobium (Nb) electrodes, and an SiO$_2$ dielectric film isolating electrodes. The design and preparation of each part are described below in detail.

\subsection{AlMn TES}\label{subsec1}

\begin{figure}[h]%
    \centering
    \includegraphics[width=0.6\textwidth]{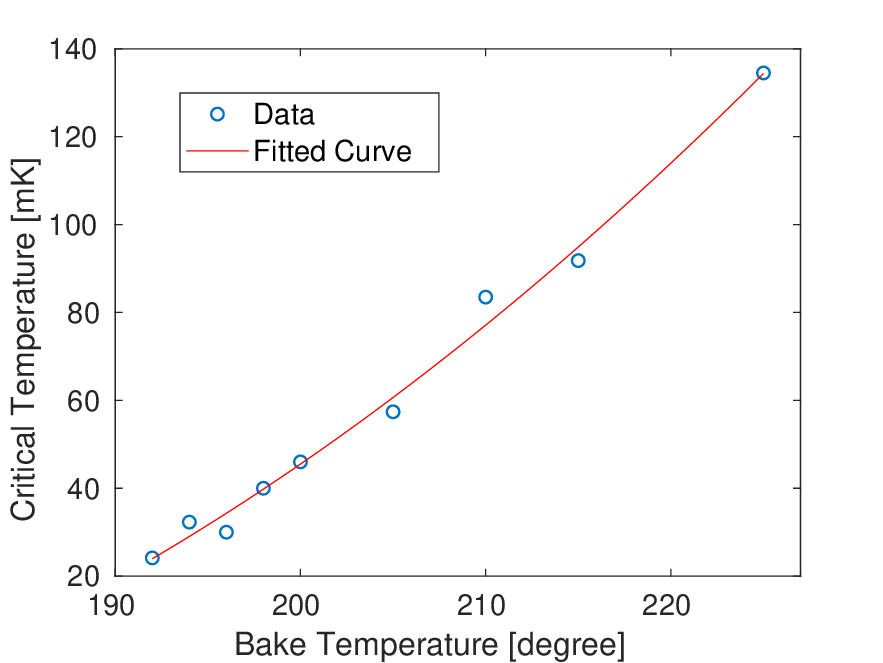}
    \caption{Relationship between critical temperature of AlMn film and baking temperature. Under atmospheric pressure, each AlMn film is thermally annealed on a hot plate for 10 minutes}
    \label{fig2}
\end{figure}

The process began by etching the backside Si$_3$N$_4$ and SiO$_2$ layers to expose the underlying silicon substrate, as shown in Fig. \ref{fig3} step (1). The remaining Si$_3$N$_4$/SiO$_2$ bilayer film served as an effective etch mask during DRIE, with sufficient thickness to withstand the plasma chemistry until complete through-wafer etching of the 200 $\mu$m silicon layer was achieved in step (10). In step (2), a 180 nm thick AlMn film was deposited via DC magnetron sputtering using an AlMn target containing 2000 ppm Mn atoms. The deposition was performed at 100 W DC power under 5 mTorr argon atmosphere. Subsequently, the film was annealed at 220 degree for 10 minutes to obtain a critical temperature (Tc) between 100 mK and 120 mK, as determined by the fitted curve in Fig. \ref{fig2}. Finally, the AlMn film was then patterned into a circle with a diameter of 100 $\mu$m using wet etching with an aluminum-specific etchant. In step (3), a 220 nm thick SiO$_2$ film was deposited by RF sputtering to partially cover the AlMn film while leaving its edge and central area exposed for electrical connection to the 240 nm thick Nb electrodes subsequently prepared in step (4). The right panel of Fig. \ref{fig1} shows the detailed layout diagram. The SiO$_2$ film serve as dual purposes: (i) as a protective layer to prevent corrosion of the AlMn film; and (ii) as an electrical insulation layer to enable proper routing of the inner Nb electrode. In step (5), a 10 nm Ti/200 nm Au bilayer film was patterned via electron-beam evaporation onto the central region of the AlMn film to provide corrosion protection during subsequent electroplating process in step (7).

The effective region of AlMn film as TES was a ring with an inner radius $R_i$ of 22.5 $\mu$m and an outer radius $R_o$ of 45 $\mu$m. Based on our measured block resistivity $\rho$ of 3.2 $\mu\Omega$·cm, the normal resistance of AlMn TES was estimated to be 19.7 m$\Omega$ by the expression $R_n=\rho$·log($R_o$/$R_i$)/(2$\pi$$t$), in which $t$ is the thickness of the AlMn film. At the operating temperature around 100 mK, radiative ballistic phonon transport will play an important role in silicon nitride film, and the heat flow from the TES to the heat bath can be expressed as $G=4\xi A\sigma T^3$, in which $\xi$ is the transport effciency, $A$ is the phonon-emitting area,  $T$ is the TES temperature, and $\sigma=157$ $W/m^2/K^4$ stands for the Stefan Boltzmann constant given by Reference \citep{Holmes-1998}. The parameter $A$ is the product of the outer perimeter of the TES and the membrane thickness. In other words, the thermal conductance of the membrane is proportional to the outer perimeter of the TES, which has been experimentally confirmed in the literature \citep{Hoevers-2005, Hays-Wehle-2016}. Once the thermal conductance (corresponding to the outer diameter) is fixed, the normal resistance can be freely adjusted by the inner diameter. This is an advantage of this design.

\subsection{Absorber}\label{subsec2}

\begin{figure}[h]%
    \centering
    \includegraphics[width=1\textwidth]{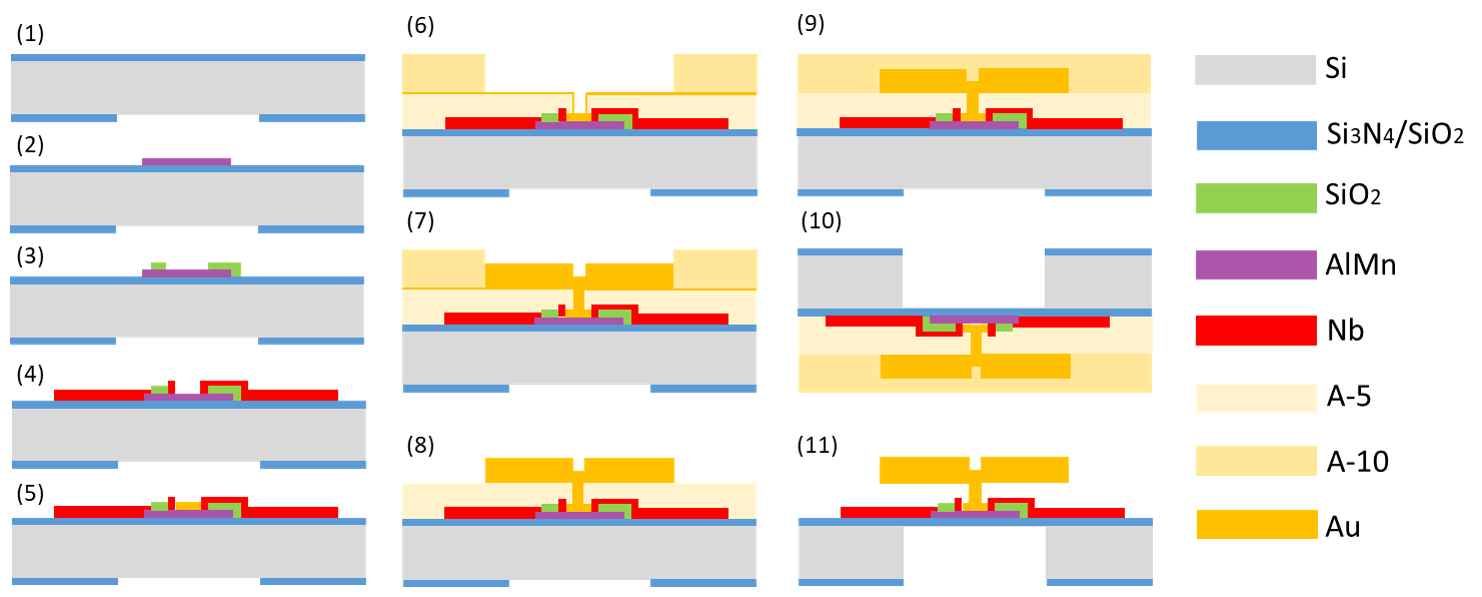}
    \caption{Schematic diagram of TES detector preparation process. (1) Etching the back of silicon nitride/silicon dioxide, (2) wet etching of AlMn film, (3) preparation of isolation layer, (4) sputtering Nb electrodes, (5) sputtering a Ti/Au film on TES, (6) defining plating zone, (7) electroplating to form an absorber, (8) acetone dissolving photoresist A-10, and etching Ti/Au seed layer, (9) using photoresist A-10 to protect the absorber, (10) making DRIE to release the membrane, (11) cleaning all photoresists to release the absorber.}
    \label{fig3}
\end{figure}

Gold or Gold/Bismuth absorbers are commonly used in the detection of soft X-rays. In this study, Au absorber was chosen just for simplification. We designed several square absorbers with side lengths ranging from 100 to 240 $\mu$m to study the influence of total heat capacity on the performance of the detector. The thickness is 2.7 $\mu$m, and the detection efficiency for 5.9 keV photons can reach 90\% in theory. The fabrication processes of Au absorber are shown in Fig. \ref{fig3}. In step (6), the silicon wafer was coated with AZ12XT-20PL-5 (A-5 for short) photoresist to lithographically define the stem, which was followed by a 20 nm / 200 nm thick Ti/Au seed layer. The photoresist AZ12XT-20PL-10 (A-10 for short) with a height of about 10 $\mu$m was used to define the electroplating area in step (7). The curing temperature of A-10 was 20 degree lower than that of A-5, to avoid the seed layer deformation caused by the solvent volatilization in A-5. Electroplating was carried out in National Institute of Metrology, China.

\subsection{Membrane}\label{subsec3}

\begin{figure}[h]%
    \centering
    \includegraphics[width=0.5\textwidth]{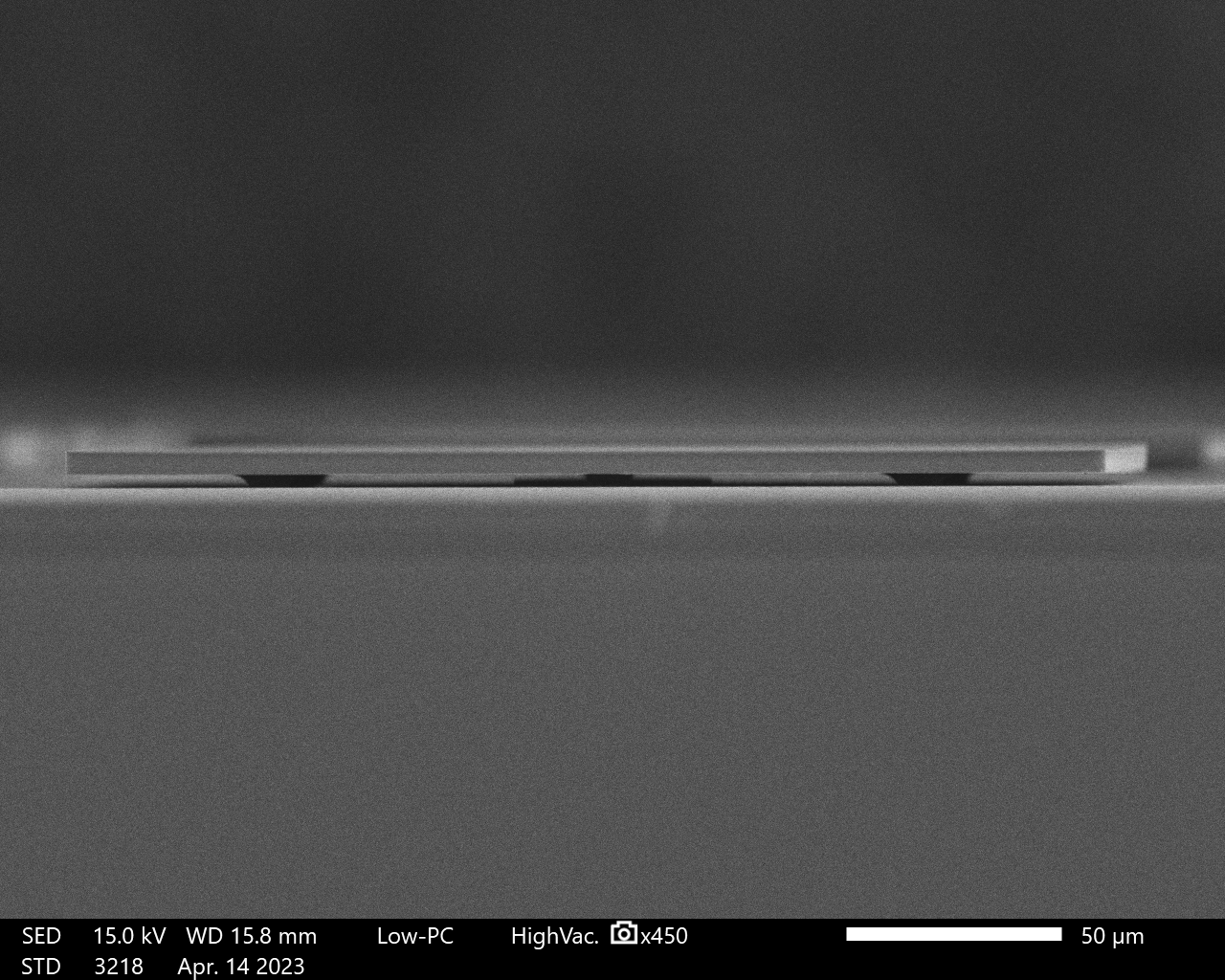}
    \caption{A side view of a TES detector. The gold absorber is square (240 $\mu$m x 240 $\mu$m) with a thickness of 2.7 $\mu$m. It is elevated about 2.5 $\mu$m from the substrate by five 10 $\mu$m-width square stems. The center stem doubles as a thermal link to the AlMn TES.}
    \label{fig4}
\end{figure}

The membrane was a Si$_3$N$_4$ / SiO$_2$ film with thickness of 1 $\mu$m / 0.2 $\mu$m respectively. It was square and the side length was 160 $\mu$m. Before releasing the membrane, two steps had been done. First, the Ti/Au seed sputtered in step (6) was etched in step (8) by Ar ions. Second, photoresist A-10 was spin-coated on the silicon to protect the absorber in step (9). Then, the 4-inch silicon wafer was inverted and glued to an 8-inch wafer tray by hot-melt glue Crystalbond 509 under 120 degree in step (10). All of them were placed in an etcher to remove the silicon through the Bosch process. Finally, the whole tray was soaked in acetone to remove the photoresist and hot-melt glue to release the absorber in step (11). Fig. \ref{fig4} shows an SEM photo of one well-made TES detector. It is clear that Au Absorber stands well on the membrane through five 10 $\mu$m width stems. There is no visual sign of collapse or deformation.

\section{Characterization and discussion}\label{sec3}

\begin{figure}[h]%
    \centering
    \includegraphics[width=1\textwidth]{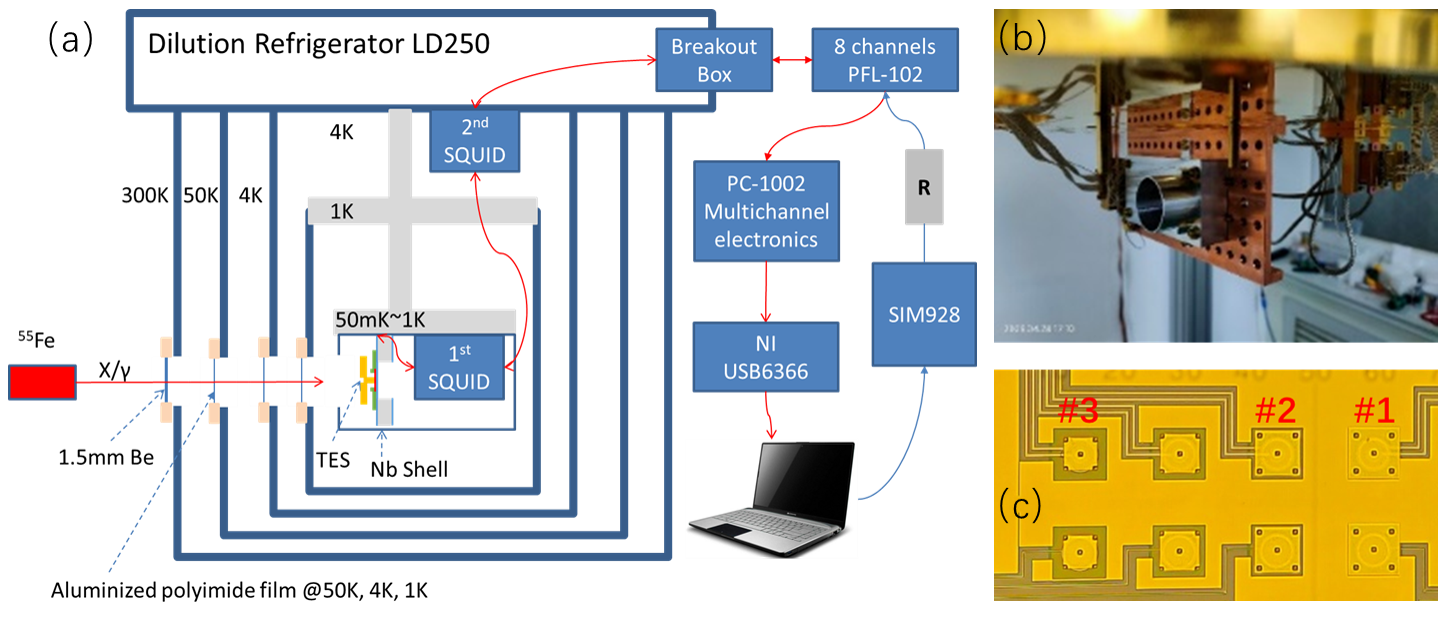}
    \caption{TES detector test system and internal physical diagram. (a) depicts the structure of the whole test bench, including a $^{55}$Fe isotope, a TES test box, a dilution Refrigerator LD250 and a two-stage DC-SQUID read out electronics. (b) shows a photo of the TES detector test box, which is thermally connected to the mixing chamber. (c) shows eight different TES detectors, among which three detectors marked as $\#1$, $\#2$ and $\#3$ are tested here.}
    \label{fig5}
\end{figure}

In order to characterize their performances and make comparison, three randomly selected TES detectors shown in Fig. \ref{fig5} (c) were tested. The side lengths of detector \#1 to \#3 are 160 $\mu m$, 140 $\mu m$, and 100 $\mu m$, respectively. They were fixed via GE Vanish glue on a copper bench, and shielded by an niobium box with a window in front of the TES. All of them were installed on the cold plate in a dilution refrigerator (DR) as shown in Fig. \ref{fig5} (b). Fig. \ref{fig5} (a) depicts the structure of the whole test bench. A $^{55}$Fe radioactive source, with an activity of about 0.7 $mCi$, was placed outside the window of the DR and about 10 cm away from the detector surface. A two-stage SQUID amplifier and corresponding room temperature electronics, produced by STAR Cryoelectronics, were used to read out the current pulse with a gain of 10$^5$ V/A. A 16-bit NI USB6366 board recorded pulses from -1 V to 1 V at a sampling rate of 2 MHz. A DC voltage source SIM928 in series with a 1000 $\Omega$ resistor supplied the required bias current. The shunt resistance was 0.3 $m \Omega$ and the inductance in series with the TES was 30 $nH$.

\begin{figure}[h]%
    \centering
    \includegraphics[width=0.6\textwidth]{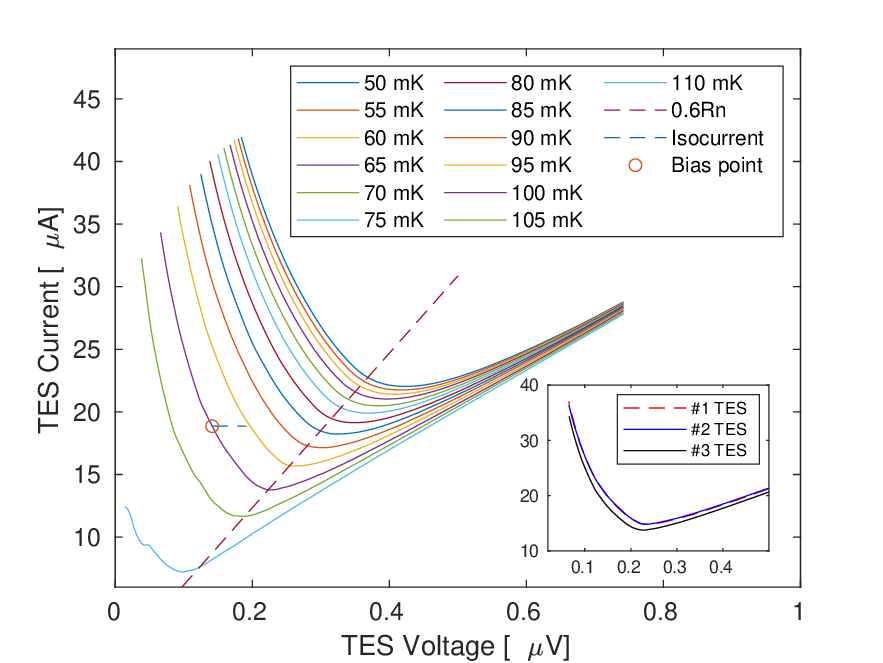}
    \caption{I-V curves of TES detector \#3 measured at bath temperatures from 50 mK to 110 mK. The inset image shows one I-V curve of each detector at 100 mK.}
    \label{fig6}
\end{figure}

\begin{figure}[h]%
    \centering
    \includegraphics[width=0.6\textwidth]{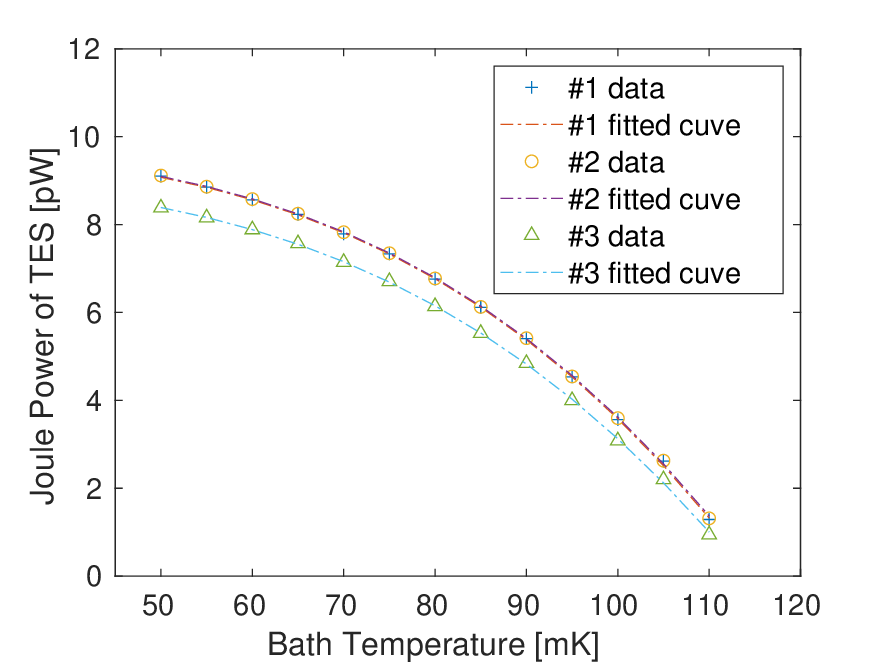}
    \caption{Joule powers of three detectors at different heat bath temperature}
    \label{fig7}
\end{figure}

Fig. $\ref{fig6}$ shows the I-V curves of detector $\#3$ tested in the bath temperature range from 50 mK to 120 mK. All detectors exhibit similar I-V characteristics, with detectors $\#1$ and $\#2$ showing nearly identical curves, as illustrated in the inset of Fig. $\ref{fig6}$. The normal resistances of three detectors were about 27 m$\Omega$, higher than the designed value of 19.7 m$\Omega$. This discrepancy may originate from deviation in film thickness and the unaccounted AlMn layer beneath the Nb electrodes. Joule powers of TES detectors working at the 60\% normal resistance were extracted and shown in Fig. $\ref{fig7}$. They were fitted by the heat flow equation $P(T)=K(T^n-T_b^n)$, in which $P=VI$ or $I^2R$ is the Joule power of TES resistance, $K$ is the prefactor, $T$ is the TES temperature, $T_b$ is the bath temperature and $n$ the index. Those parameters including thermal conductance $G=nK^{(n-1)}$ are summarized in table \ref{table2}. All three detectors have almost the same characteristics. The critical temperatures are identical with the expected value of 116 mK given by the red line in Fig. $\ref{fig2}$. A thermal index between 3.2-3.3 maybe indicates the coexistence of radiative and diffusive phonon transport mechanisms in the silicon nitride film. At the critical temperature of 114.1 mK, the thermal conductance of detector $\#3$ is about 255 pW/K, slightly lower than the design value of 293 pW/K given by function of $G=4\xi A\sigma T^3$, where we assume that $\xi$ is equal to 1 and the thickness of membrane without SiO$_2$ film is about 1 $\mu$m.

\begin{figure}[h]%
    \centering
    \includegraphics[width=0.6\textwidth]{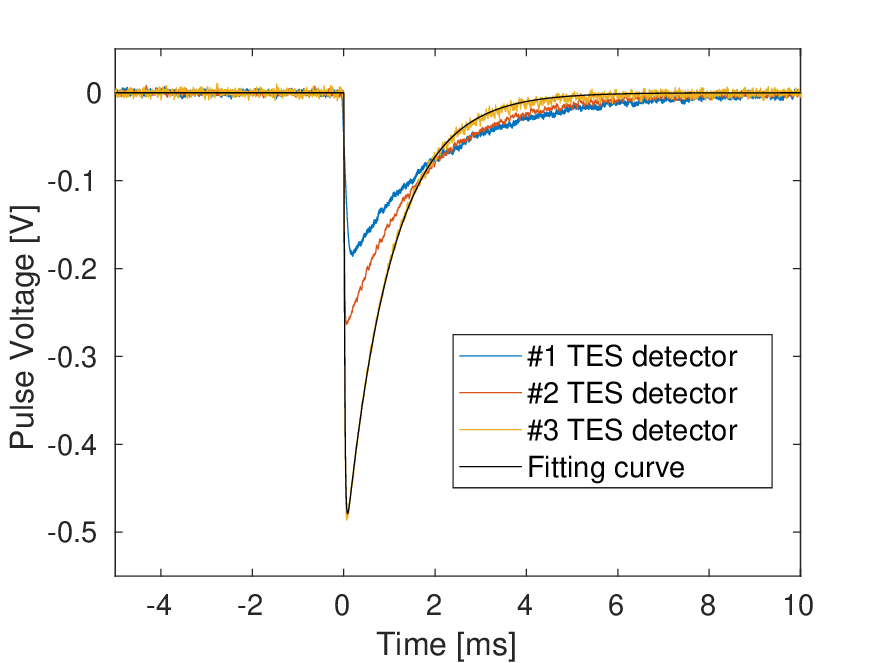}
    \caption{Response pulse of TES detector to 5.9 keV photon}
    \label{fig8}
\end{figure}

\begin{table}[h]
\caption{Characteristics of three TES detectors}\label{tab2}%
\begin{tabular}{@{}lllllll@{}}
\toprule
Parameter & Illustration & Detector \#1 & Detector \#2  & Detector \#3 \\
\midrule
Absorber [$\mu$m$^3$] & & 160$\times$160$\times$2.7  & 140$\times$140$\times$2.7 & 100$\times$100$\times$2.7 \\
$T_c$ [mK] & Critical temperature & 115.0 & 115.2 & 114.1 \\
$K$ [pW/K$^n$] & Prefactor & 1.22$\times$10$^{-8}$ & 1.19$\times$10$^{-8}$ & 1.00$\times$10$^{-8}$ \\
$n$ & Index & 3.30 & 3.29 & 3.23 \\
$G(Tc)$ [pW/K] & $G(Tc)=nK{T_c}^{n-1}$ & 278 & 278 & 255 \\
$T_0$ [mK]& & 112.6 & 112.7 & 112.3 \\
$I_0$ [$\mu$A] & & 20.8 & 20.9 & 18.7 \\
$R_0$ [m$\Omega$] & & 6.77 & 6.74 & 7.49 \\
$P_0$ [pW] & $P_0={I_0}^2R_0$ & 2.93 & 2.94 & 2.67 \\
$G_0$ [pW/K] & $G_0=G(T_c)({\frac{T_0}{T_c}})^{n-1}$& 265 & 264 & 246 \\
$\alpha_I$  & $\alpha_I=\frac{T_0}{R_0}\frac{\partial R}{\partial T}|_{I_0}$ & 75 & 67 & 73 \\
$\beta$$_I$ & $\beta_I=\frac{I_0}{R_0}\frac{\partial R}{\partial I}|_{T_0}$ & 2.6 & 2.1 & 1.9 \\
$L_I$ & $L_I=\frac{\alpha_IP_0}{G_0T_0}$ & 7.4 & 6.8 & 7.0 \\
$\tau_{-}$ [ms] & & 2.1 & 1.6 & 1.0 \\ 
$\tau$ [ms] & $^{(1)}$$\tau \approx \tau_{-}\frac{1+\beta_I+L_I}{1+\beta_I}$ & 6.5 & 5.1 & 3.4 \\
$C_0$ [pJ/K] & $C_0=\tau G_0$ & 1.72 & 1.33 & 0.85 \\
Estimated FWHM & & 5.0 & 4.4 & 3.5 \\
@ 5.9 keV [eV] \\
Tested FWHM & & 20.5$\pm$1.2 & 15.3$\pm$1.1 & 11.0$\pm$1.0 \\
@ 5.9 keV [eV] \\
\botrule
\end{tabular}
$^{(1)}$: The inductance in series with TES was 30 $nH$. It was sufficiently small to make the function feasible. 

\label{table2}
\end{table}

The performance calibration of TES detectors was carried out at an operating point: the heat bath temperature was 100 mK and the bias current was 0.49 mA. The corresponding parameters including TES temperature $T_0$, TES current $I_0$, and TES resistance $R_0$ were extracted from I-V curves and listed in table \ref{table2}. Fig. \ref{fig8} shows the 5.9 keV photon signal pulse of each detector. Detector $\#$3 gets the maximum pulse because of its smallest absorber. We extracted the pulse peak value and decay time $\tau_-$ by fitting the data with the double exponential function from Reference \citep{Irwin-2005}. The temperature sensitivity $\alpha_I$ and current sensitivity $\beta_I$ were determined from derivatives of the TES resistance with respect to temperature (along isocurrent lines in Fig. \ref{fig6}) and current (along isothermal lines), respectively. Then, the loop gain $L_I$ was further derived. The derived values $\alpha_I$, $\beta_I$, and $L_I$ were used to deduce the heat capacities $C_0$ of the TES detectors according to the relationships shown in Table \ref{table2}. Measured heat capacity values are about 3-4 times greater than the expectation given by Reference \citep{Brown-2008}. The observed enhancement in the heat capacity of the TES detector currently lacks a definitive explanation. To systematically investigate this discrepancy, further experimental and analytical work, such as testing TES detector without the Au absorber to exclude potential contributions from AlMn TES, is required in the future to help clarify whether the enhancement arises from the absorber, the TES film, or an unanticipated systematic effect in the measurement.

\begin{figure}[h]%
    \centering
    \includegraphics[width=0.8\textwidth]{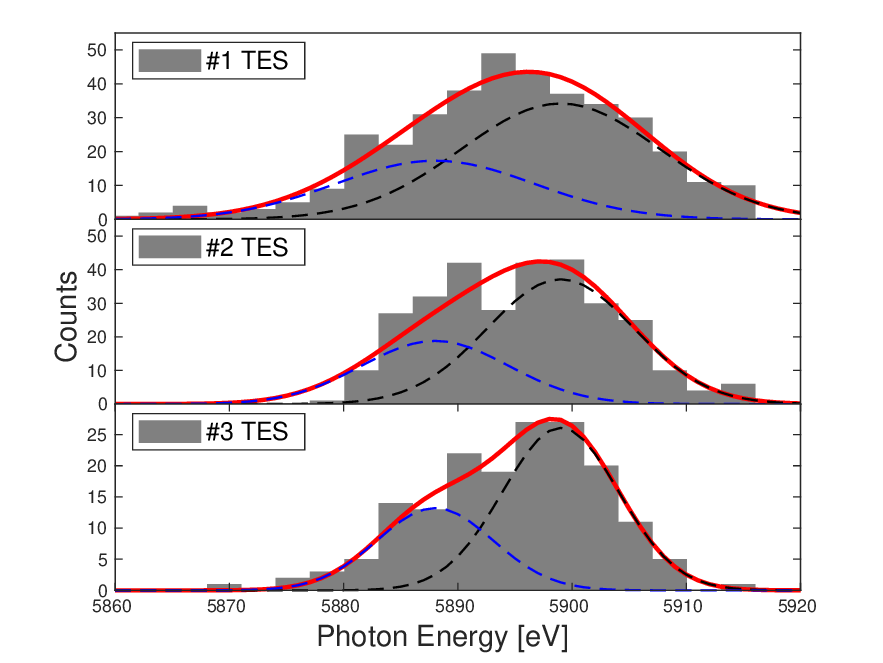}
    \caption[width=0.8\textwidth]{Energy spectra of TES detectors. A double-gauss function (red line) was used to fit the experimental data. Dark dashed line stands for the component of K$_{a1}$ X-ray, and Blue dashed line for K$_{a2}$ X-ray.}
    \label{fig9}
\end{figure}

\begin{figure}[h]%
    \centering
    \includegraphics[width=0.8\textwidth]{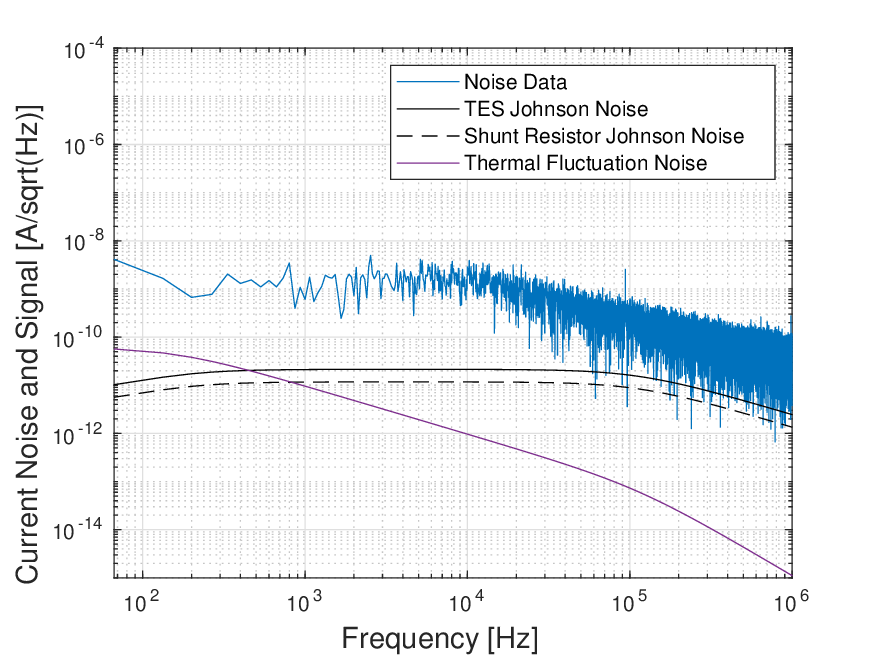}
    \caption{The noise spectra of detector \#3}
    \label{fig10}
\end{figure}

Fig. \ref{fig9} shows the statistic results of pulse peak values coming from fitting. Due to the low activity of radioactive source, the effective time loss caused by SQUID jumping, and the noise change due to external interference, the total effective counts of the three detectors are 376 ($\#$1), 300 ($\#$2) and 171 ($\#$3), respectively. Fitting results show that the energy resolution of detector $\#$1 is 20.5 $\pm$ 1.3 eV @ 5.9 keV, while that of detector $\#$2 is 15.3 $\pm$ 1.1 eV and detector $\#$3 is 11.0 $\pm$ 1.0 eV. In order to make comparison between the measured energy resolution and the theoretical value, we calculated the Johnson noise and thermal fluctuation noise utilizing the functions provided in Table 1 of the Reference \citep{Irwin-2005}, and further obtained the energy resolution of each detector. The function $F(T_0,T_b)$ appeared in the noise calculation is about 0.8 both at the radiative limit and the diffusive limit of the membrane. As shown in Fig. \ref{fig10}, the average current noise below 20 kHz is approximately 1.68 nA/$\sqrt{Hz}$, which is significantly larger than the combined contribution from the Johnson noise of the TES and shunt resistors and the thermal fluctuation noise. Those calculated noises account for only 3.5 eV, which means that other noises including SQUID noise and excess noise constitute the majority of the 11.0 eV. Other detectors have the same situation, as shown in Table \ref{table2}. Therefore, reducing noise will result in better resolution.

\section{Conclusion}\label{sec4}

In this study, we deigned and fabricated samples of X-ray detectors based on annular AlMn TES. This design would theoretically allow us to easily adjust the normal resistance of TES without changing its thermal conductivity to the heat bath. Experimental demonstration will be performed in the next study by fabricating and testing multiple TES devices with varying inner-to-outer radius ratios. The SEM photo shows that the absorber stands well on the membrane through five stems, with no visual sign of collapse or deformation. The I-V curves of the three TES detectors have good consistency. Experimental results show that the total heat capacities of TES detectors are about 3-4 times greater than expected. Heat capacity is an important parameter and any deviation will affect the final performance of the TES detector. In the future, we will fabricate new TES detectors and conduct more detailed tests on them to clarify the source of the excess capacity. Among those TES detectors, Detector \#3 has the best resolution of 11.0 eV @ 5.9 keV, which is significantly greater than the estimated value of 3.5 eV. The analysis of noise spectra indicates that there are some noise components larger than expected in the current test system. The reduction of this noise will lead to better performance in the future. 

This study once again illustrates the application of AlMn TES in X-ray detection. When applying it to gamma ray detection, we can enlarge the absorber size to the level of millimeter to enhance the detection efficiency, coupled with a larger annular TES sensor to increase the thermal conductivity. Meanwhile, the normal resistance can be easily controlled by tuning the inner-to-outer radius ratio. The baking method can also be inherited to adjust the critical temperature. In combination with its relative simplicity in array fabrication, it is a promising solution for future space X-ray and gamma-ray missions with large-scale detector arrays.

\backmatter

\bmhead*{Acknowledgments}

This work is supported by the National Natural Science Foundation of China (Grant Nos. 12275292, 12141502) and the National Key Research and Development Program of China (Grant No. 2021YFC2203402).

\bibliography{sn-bibliography}

\end{document}